# Fundamental Fracture Mechanics Equation of material

L.S.Kremnev

*Moscow State Technological University «Stankin», Moscow, Russia*
kremnevls@yandex.ru

Abstract: On the basis of energy conservation law an without utilizing Linear Fracture Mechanics (LFM) postulates the equation of a real-structure material elastic-plastic fracture has been derived. With the help of this equation the force and energy criteria of Non-linear Fracture Mechanics (NLFM) have been found. These criteria constitute the basis of modern strength analysis of machine-parts and structures made of real-structure materials. $K_{1c}$ dependence on ultimate strength limit, yield limit and impact toughness has been established and experimentally confirmed for a number of steels.

It is considered that Elastic Fracture Mechanics (EFA) provides the basis for modern strength analysis of bearing structures and parts. Equation (I) for force criterion ($K_{1c}$) and equation (II) for energy criterion ($G_{1c}$) of material fracture are referred to as the fundamental equations of EFA, these criteria being equal to:

$$K_{1c} = c\sigma\sqrt{\pi\, l} \qquad (I)$$

$$G_{1c} = K_{1c}^2 / E \qquad (II)$$

where $K_{1c}$ – the Critical Stress Intensity Factor (CSIF), crack resistance (CR), $\sigma$ - ultimate stress limit of a specimen with a crack with length $l$, $c$ – a factor, dependent on a specimen form and a test conditions, $G_{1c}$ –crack toughness, $E$ – normal modulus of elasticity.

The abovementioned fracture criteria were obtained with the following assumptions taken into consideration:

1. The material being fractured was elastic, i.e. its relative strain is directly proportional to the stress applied. Therefore the fracture mechanics based on this assumption is called Elastic Fracture Mechanics «EFM».

2. The material structure is continued (in contrast to discrete i.e. atomic structure).

3. The crack tip radius $\rho \rightarrow 0$, and the stress in its corner $\sigma_y \rightarrow \infty$.

The assumptions mentioned make up the physical model of fracturing, on the basis of which its force and energy EFM criteria can be obtained. Nevertheless this model is rather far from real conditions of operation and research of the parts made of these materials (as well as of their fracture patterns)

In addition to these, the dimension of $K_{1c}$ [ $Pa\,m^{1/2}$ ], in contrast to the dimensions of other structural and tool material properties' values, fails to adequately reflect its substantial content. In this regard it is feasible to mention the words of J.F. Nott [1], a prominent specialist in the field of Fracture Mechanics: "The physical sense of parameter $K$ is difficult to understand mostly due to its dimension ( *stress × square root of length*), which is rather hard to physically imagine"

Let's consider the plate with thickness $\delta$ made of a material which is not necessarily elastic. The plate dimensions in the drawing plane are not confined and the plate is affected by the stress $\sigma$ (Fig.1). Let's assume that in a certain area of the plate there appeared a stress $\sigma_y > \sigma$ due to which a through crack was formed. The crack length is *2l* and its width is *2a* (Fig.1). The task consists in determining the relationship $\sigma_y = \varphi(\sigma, l, a)$ for the system equilibrium condition.

The direct cause of the appearance of the *2l* - long crack is the presence of stress $\sigma_y$ being formed in its corners, whereas the mediate cause of the crack is in the action of stress $\sigma$ applied to the plate (Fig.1). It is evident that the fracture energies $W_y$ and $W$ of forces $P_y$ and $P$ which account for the appearance of stresses $\sigma_y$ and $\sigma$ are equal. Let's determine the values of these energies and set them equal:

$$W_y = P_y \times S_y \quad (1)$$

where $P_y$ – fracturing force, $S_y$ –transfer(broadening) of a crack in the direction of $P_y$. In its turn (Fig.2):

$$P_y = (\sigma_y - \sigma) \times \delta \times \Delta, \quad (2),$$

where $(\sigma_y - \sigma)$ – ultimate stress of the plate which is affected by the action of stress $\sigma$ before the appearance of $\sigma_y$ stress (actually if $\sigma_y = \sigma$, fracture force $P_y = 0$, and the plate remains undisturbed by definition, but under action of the stress $\sigma$); $\Delta$ -the minimal distance of the crack edges irretrievable displacement appearing in its tip under the action of ultimate stress (Fig.2). It is evident that in a material with real discrete structure the value $\Delta$ is equal to interatomic spacer $b$. Thus, the crack opening displacement energy is:

$$W_y = \frac{1}{2}(\sigma_y - \sigma) \times \delta \times \Delta \times 2a \quad (3)$$

Here the cofactor $2a$ (crack opening) represents the displacement of crack $S_y$ in the direction of force $P_y$ taking place during crack extension up to the length $2l$; the crack opening value $2a$ is a sum of irretrievable displacements $\Delta$ of both crack edges taking place at its tip ($2a = \Sigma \Delta$); the cofactor ½ takes into account the fact that the ultimate stress varies from 0 to $(\sigma_y - \sigma)$. The work of force $P$ which is necessary to create those two crack faces (Fig. 2) equals to:

$$W = \sigma \times 2l \times \delta \times \Delta \quad (4)$$

Having set the equations (3) and (4) equal, we obtain:

$$\text{½} (\sigma_y - \sigma) \times 2a = 2\sigma l, \quad (5)$$

From equation (5) we can get the desirable equation:

$$\sigma_y = \sigma(1 + 2l/a) \quad (6)$$

In line with (6) for a plate having a circular hole with radius $r$ ($r = l = a$) we obtain:

$$\sigma_y = 3\sigma \quad (7)$$

If the crack is assumed to have an elliptical form then from equation (6) (using its property $\rho = a^2/l$) we obtain:

$$\sigma_y = \sigma (1 + 2\sqrt{l/\rho}) \quad (8)$$

As it is known from [2] this equation is correct for a crack of any form on the periphery of which there exists a point with a small curvature radius $\rho$. Taking into consideration the fact that $2\sqrt{l/\rho} \gg 1$ we can present (8) in the following form:

$$\sigma_y = 2\sigma\sqrt{l/\rho} \quad (9)$$

This equation is also valid for a lateral crack with length $l$ and curvature radius $\rho$. Let's multiply both parts of (9) by $\sqrt{\rho}$:

$$\sigma_y \sqrt{\rho} = 2\sigma\sqrt{l} \quad (10)$$

In case of fracturing the plate having an edge crack with the length l the right-hand part of (10) can be transformed into the force criterion of fracture mechanics $K_{Ic}$, i.e. into equation (I).

$$\sigma_y \sqrt{\rho} = K_{1c} = 2\sigma\sqrt{l} = 1{,}13\,\sigma\sqrt{\pi l} \tag{11}$$

where $\sigma = \sigma_в$ – ultimate stress limit [the external stresss in contrast to the internal one $\sigma_y - \sigma$ used in equation (2). The equations (6) –(9) are well known. They were obtained by G. Kolosov, K.Inglis, A.Griffits, and the right-hand part of equation type (11) can be found in the researches of J.Irvin for the case of material fracturing with the abovementioned assumptions 1-3 taken into account, that is for the elastic continuum of ideal structure having a crack with $\rho \to 0$, and $\sigma_y \to \infty$. Let's prove that the main equation (5) and subsequent equation (11) are obtained on the basis of the energy conservation law without the assistance of EFA postulate concerning the linear dependence of relative material deformation on the applies stress. Let's write equation (5) for a plate with a lateral crack, having the length l, at the tip of which the ultimate stress $\sigma_y - \sigma$ (Fig.3) is observed.

$$(\sigma_y - \sigma) \times a = 2\sigma l, \tag{12}$$

Here a – lateral crack opening, which is twice as small as that of central crack because it is equal to $\Sigma\Delta$ as a result of diverging the faces of one of the crack tips, in contrast to the two crack tips diverging as it takes place in case of a central crack (Fig.1,3). The left-hand part of equation (12) represents the energy necessary to create a unit fracture surface area during the crack extension i.e. fracture toughnes $G_{1c}$ [J/m$^2$]. Similar physical significance and dimension can be also attributed to $J_{1c}$ – integral obtained by J. Rise [3] on the basis of energy conservation law as an energy criterion of elastic-plastic material fracture (nonlinear fracture mechanics criterion).

$$J_{1c} = \sigma_0 \times \delta_к, \tag{13}$$

where $\sigma_0$ –breaking stress at a lateral crack tip, $\delta_к$ – lateral crack opening ($\delta_к = a$). Along with these the multiplicands of the right-hand part of equation (12) $\sigma$ and $l$ unambiguously determine the value $K_{1c}$= 1.13 $\sigma\sqrt{\pi l} = 2\sigma\sqrt{l}$ (11). By means of some simple reexpressions we can come up with the energy criterion of nonlinear fracture mechanics of a real-structure material (NEFM):

$$G_{1c} = J_{1c} = (K_{1c})^2 / 2\sigma_в \tag{14}$$

From (14) we can find: $K_{1c}(CR) = \sqrt{2\sigma_в \times G_{1c}(J_{1c})}$, which was to be proved.

One should also pay attention to the fact that equations (5) and (11) were obtained in this work without applying Hooke Law.

At the same time of the fracture process consideration on the basis of the presented crack model (Fig.1,2) makes it possible to exclude the assumption consisting in the fact that the radius of a crack tip curving in the plate made of a real-structure material with discrete structure $\rho \to 0$, and the stress in its tip $\sigma_y \to \infty$. Actually $\rho$ in such material cannot be less than interatomic space b and $\sigma_y$ cannot exceed the theoretical ultimate limit $\sigma_{theor}$. As a consequence of this fact it occurred that the cracking resistance of material with a blunt crack notably exceeds (by 13%) that with a sharp crack: $1.13\sigma\sqrt{\pi l}$ (equation (11)) and $\sigma\sqrt{\pi l}$ [2] respectively). Evidently it should bee taken into account that the values $K_{1c}$ provided in reference literature usually refer to the cracks with $\rho \to 0$.

The universal, two-fold character of the force fracture criterion $K_{1c}$ is reflected in equation (11). In its left-hand part the stress intensity factor in case of brittle, that is elastic fracture $K_{1c}(CSIF)$ is equal to $\sigma_y\sqrt{\rho}$, and the one in its right-hand side – to cracking resistance $K_{1c}(CR) = 2\sigma\sqrt{l}$ for elastic-plastic fracturing.

In the first case $K_{1c}(CSIF)$ is a force criterion of EFM because its value only depends on $\sigma_y \approx 0.1E$ [4] and $\rho \approx b$, that is on the properties of material with ideal structure. Such material doesn't contain any defects (in the first place dislocations the migration and generation of which are responsible for material plastic deformation) and thus it is subject to elastic, brittle fracturing. The specimens made of defect-free materials have been obtained in many laboratories. Their experimental ultimate strength $\sigma_в$ appeared to be equal to the relevant calculated value $\sigma_{theor.} = \psi(E,b)$. From the left-hand part of (11) it follows that the iron-base alloys have $K_{1c}$ (CSIF) ≈ 0,38 МПам$^{1/2}$ (E ≈ 220 000МПа, b≈0.3 10$^{-10}$ м). Nevertheless the value $K_{1c}(CR)$ in steels (the alloys made on the basis of iron) calculated in accordance with the right-hand part of (11) with the experimental values σ and l taken into consideration i.e. for the case of elastic-plastic material fracturing, considerably exceeds the values of $K_{1c}$ (CSIF) and amount to 100 – 130 МПам$^{1/2}$, that is 250 – 300 times as large. These data confirm the conclusions drawn by E. Orovan [5] who experimentally demonstrated that the values of plastic fracturing energy of are several orders of magnitude higher than those of brittle fracturing energy. Consequently the criteria $K_{1c}$ (CR) and $K_{1c}$(CSIF) do not coincide either by their sense nor by their values and one has to distinguish them.

At the same time it turned out that the criterion $K_{1c}$ (equation I), obtained by D.Irvin on the basis of a fracturing model and in accordance with the assumptions (1-3) EFM can be used as a criterion of elastic-plastic fracturing of a real-structure material. It was demonstrated that the acceptable values of crack lengths $l_{cr}$ of different materials under the plain-strain conditions calculated in accordance with (I) satisfactorily coincide with the relevant experimental results. This unexpected result was explained by D. Irvin by the fact that at the crack tip there appears a thin layer of plastically deformed material in the presence of which the latter transforms into quasi-brittle condition and therefore the asymptotic equations EFM are still valid. Usually the size of this thin layer is confined – it is considered to be no more than 20% of a crack length [6]. Nevertheless $K_{1c}$ value (equation I ) is widely and successfully used for determining the critical crack length $l_{cr}$ of numerous constructional materials, in which the process of a crack forming is preceded by forming the sizable zone of plastically deformed material. To these materials one can attribute e.g. high-temper structural steel*s*. The linear dimension of plastic deformation zone ▼ at the lateral crack tip taking place immediately prior to its expansion under the plain-strain condition is equal to [3]:

$$▼ = (1/3π) (K_{1c}/σ_T)^2 \qquad (15)$$

From equation (11) we can obtain:

$$L_{cr.} = 0,25 (K_{1c}/σ_в)^2 \qquad (16)$$

Using (15) and (16) one can find relative dimension of plastic deformation zone at the edge crack tip just before the material fracture:

$$▼ / l_{cr.} = 0,42 (σ_в/σ_T)^2 \qquad (17)$$

By means of equation (17) one can estimate the relation ▼ / $l_{cr.}$. Referring to a typical structural steel grade 40X. *Fter* its standard heat treatment the steel hardness is as high as 30 – 32HRC, $σ_в$ and $σ_т$ are respectively equal to 1030MPa и 800MPa [7]. In accordance with equation (17) the length of plastic zone at the crack tip of steel 40X comprises 70% of its total length. In steel 3 the size of this experimentally determined zone [6] considerably exceeds the crack length.

The abovementioned values of ▼/$l_{кр}$ substantially exceed the value of a thin layer, this zone as it was mentioned shouldn't be larger than 20% $l_{cr}$. Thus the explanation of the fact that the laws of EFM hold true for elastic-plastic materials due to their transferring into quasi- brittle condition because of creating a thin deformed layer at the crack tip do not seem to be convincing enough. With that the obtained proof of the fact that the force fracture criterion $K_{1c}(CR)$ **(I)** results from the energy conservation law without its restricting by EFM postulates provides direct explanation to the existing situation**.**

It is evident that the supposition concerning the quasi-brittle fracturing of real materials was put forward to substantiate the possibility of EFM force criterion $K_{1c}$ (CSIF) application under conditions of elastic-plastic fracture of these materials. In realty the explanation consists in the fact that $K_{1c}(CR)$ in itself is a criterion of non-linear mechanics of the real-structure materials fracture. The value $G_{1c}$ is often chosen as energy criterion of elastic-plastic fracture [3, 8] and calculated by equations (II) or (18):

$$G_{1c} = J_{1c} = [K_{1c}(CR)]^2/E \qquad (18)$$

One should mention a considerable difference between the values of $G_{1c}$, obtained by equations (14) and (18).

Equation (14) makes it possible to calculate $J_{1c}$ –integral ($G_{1c}$) by the experimental values of $K_{1c}(CR)$ and $σ_в$. Up to the recent time $J_{1c}$ was determined on the basis of specially developed and complicated experiments [3].

As equation (14) was obtained from the main equation (5) and from equation (12) which follows from it, equation (14) therefore holds true for both elastic-plastic and really brittle fracture. Let's check this statement by the example of the latter one. In this case in equation (14) instead of $K_{1c}(CR)$ one should naturally introduce $K_{1c}(CSIF)$. As a result of brittle fracture a new surface appears. To form this surface some amount of energy should be spent $E$ [J] = $γ × S$, где $γ$ [J/m$^2$] –surface energy, $S$ [м$^2$] –fracture surface area. For iron (Fe$_α$), occurring in the solid state, it has been experimentally established [9] that $γ$ = 1750 erg/cm$^2$ (1.75 J/m$^2$). The physical content of notions $γ$ and $G_{1c}$ in case of true brittle fracture is the same, because the experimental conditions (zero creeping method) on measuring the value of γ excludes plastic deformation of material. If one takes into consideration the fact that/ as it was shown earlier, in Fe$_α$ the value $K_{1c}(CSIF) ≈ 0.38$ MPam$^{½}$, and $σ_в = σ_{theor.} ≈ 22000$ MPa, then after substituting these values into (16) we can find out that for iron $G_{1c} ∼ 3.28$ J/m$^2$. Due to the fact that fracture process results in creating two surfaces then the experimental value of surface energy $γ$ and the value $G_{1c}$ determined by (14) will practically coincide: 1.75 and 1.64 J/m$^2$. The difference between these two values calculated with the help of equation (18), is as large as 500%.

In checking the justice of the above statement in case of elastic-plastic fracture one should bear in mind that the impact toughness, e.g. *KCU*, and fracture toughness $G_{1c}$ are equal to specific energy of fracture surface creation [J/m$^{2`}$]. Therefore it is feasible to compare their experimental and calculated values obtained by means of equations (14) and (18). For instance for maraging steel *H18K9M5T* $σ_в ≈$ 2000MPa, $K_{1c} ≈ 60$ MPam$^{½}$ [10]. In accordance with equations (14) and (18) $G_{1c} ≈ 0.90$MJ/m$^2$ and $≈ 0.014$ MJ/m$^2$ respectively. The experimental value $KCU ≈ 0.40$MJ/m$^2$ [11]. Thus the value $G_{1c}$, calculated by equation (14) resulting from the main equation (5), is larger whereas the same value calculated by (14) is much less than the impact strength value. If the first result can be naturally explained by the fact that *KCU* was obtained by means of notched specimens concentrating the stress in the extent directly proportional to their length, the second one can be only accounted for by the unreliability of equation (18) ( the value $K_{1c}$ doesn't depend on the crack length at a standard specimen).

Therefore equation (18) fails to be true either for true brittle or for elastic-plastic fracture, it has an eclectic character.

One should pay attention to the similarity of the conditions of fracturing the specimens with a crack induced at their tips or those of notched specimens aimed at determining their $K_{1c}$ or *KCU* because both conditions result in creating the plain-strain condition during the specimens testing [12]. Increased rate of loading (5 – 10m/sec) during impact testing in comparison with crack-resistance testing has just slightly affects the plastic properties of material. The main difference consists in the fact that the presence of a notch considerably reduces the energy of a specimen fracture whereas the length of an induced crack do not affect the value of $K_{1c}(CR)$ because the latter is a material constant. It is possible to take into consideration the influence of the notch length and its tip radius. Thus, in impact testing of a standard specimen with a *U*- notch the rupture stress limit ($σ_y$ - $σ$) at its tip increases by 2.83 times in accordance with equation (9) ($l$ =2mm, $ρ$=1mm). Bearing in mind that $G_{1c}$ = $J_{1c}$ = 2.83 × *KCU*, one can get from equation (14) the dependence of $K_{1c}$ on *KCU* and $σ_в$:

$$K_{1c} = 2{,}38 \sqrt{σ_в × KCU} \; МПам^{½} \qquad (19)$$

Along with these, in (19) it is necessary to consider the fact that in case of clear elastic-plastic fracturing the value $K_{Ic}$ in (19) shall be increased up to $K^{\blacktriangledown}_{Ic}$ value due to creating a zone of a plastically deformed material zone at the crack tip.

In accordance with J. Irvin we can write down:

$$K_{Ic}(el\text{-}pl) / K_{Ic}(el.) = \sqrt{(l + \blacktriangledown)} / \sqrt{l} \qquad (20)$$

Besides, we shall proceed from the following approximation:

$$\sigma_в \sqrt{l} = \sigma_m \sqrt{(l + \blacktriangledown)} \qquad (21)$$

From (20) and (21) we can obtain:

$$K_{Ic}(el\text{-}pl.)/ K_{Ic}(el.) = \sigma_в / \sigma_T \qquad (22)$$

Equation (19) takes on its final form:

$$K_{Ic}(CT) = 2{,}38\ \sigma_в/\sigma_T\ (\sqrt{\sigma_в} \times KCU)\ MPam^{½} \qquad (23)$$

Relationship (23) using the literary data on the values $\sigma_в$, $\sigma_m$, $KCU$ and $K_{Ic}(CR)$ for a number of steels makes it possible to obtain satisfactory results. Let's consider several typical examples (Table).

Table.

Comparison of calculated and experimental data on structural steels crack resistance $K_{Ic}(CR)^p$ and $K_{Ic}(CR)^{exp.}$.

| № | Steel grade and heat treatment mode | $\sigma_в$, MPa | $\sigma_T$ MPa | KCU, MJ/m² | $(K_{Ic})^{cal.}$ MPa × × m^{1/2} | $(K_{Ic})^{exp.}$ MPa × × m^{1/2} | $\|(K_{Ic})^{cal.} - (K_{Ic})^{exp.}\| : (K_{Ic})^{exp} \times$ × 100 %. | Reference $(K_{Ic})^{exp.}$ |
|---|---|---|---|---|---|---|---|---|
|   | 15Х2Г2НМФБ |   |   |   |   |   |   | [11] |
| 1 | Mode №1 | 1440 | 1110 | 1.00 | 117 | 130 | 10.0 |   |
| 2 | Mode №2 | 1470 | 1140 | 1.20 | 129 | 130 | 1.0 |   |
| 3 | Mode №3 | 1470 | 1250 | 1.30 | 122 | 125 | 2.4 |   |
| 4 | Mode №4 | 1480 | 1250 | 1.35 | 125 | 145 | 13.8 |   |
| 5 | Mode №5 | 1530 | 1250 | 1.20 | 125 | 120 | 4.2 |   |
| 6 | Mode №6 | 1530 | 1260 | 1.20 | 124 | 125 | 1.0 |   |
| 7 | 10Н3М3Б quenching tempering 500-550°C | 1100 | 1000 | 1.20 | 95 | 86 - 89 | 8.6 | [13] |
| 8 | 38ХС, Isotherma quenching | 1400 | 1200 | 1.20 | 114 | 113-115 | 0 |   |
| | 40X | | | | | | | [14] |
| 9 | Mode №1 | 1900 | 1580 | 0.20 | 56 | 56 | 0 | |
| 10 | Mode №2 | 1920 | 1570 | 0.30 | 70 | 68 | 3.0 | |
| 11 | Mode №3 | 1910 | 1580 | 0.28 | 66.5 | 73 | 11.6 | |
| | 45ХН2МФА | | | | | | | |
| 12 | Mode №1 | 2170 | 1820 | 0.50 | 93.5 | 46 | 103 | |
| 13 | Mode №2 | 2120 | 1680 | 0.20 | 61.8 | 64 | 3.4 | |
| 14 | Mode №3 | 2460 | 2200 (~) | 0.09 | 39.6 | 39 | 1.5 (~) | |
| | 40ХН2МФА | | | | | | | [3,15] |
| 15 | Mode №1 | 1720 | 1560 | 0.40 | 69 | 50 - 83 | 3.8 | |
| 16 | Mode №2 | 1640 | 1410 | 0.55 | 83 | 75 - 91 | 0 | |
| 17 | Н18К9М5Т | 2000 | 1900 | 0.50 | 79 | 60 | 24 | [10] |

Let us agree that in some cases when the values $(K_{Ic})^{cal.}$ and $(K_{Ic})^{exp.}$ differ more than by 25%, it is not feasible to compare them. Position №12 can be taken as an example where the steel with an ultimate stress limit 2170MPa has obviously overestimated impact strength: $KCU = 0,5MJ/m^2$ or improbably small crack resistance: $(K_{Ic})^{exp} = 46$ MPa×m$^{1/2}$.

Apparently it can be in the first place explained by insufficiently reliable results of determining the impact strength. It is conditioned in much by the absence of standardized technique for making impact test specimens (and in the first place – forming the notches on their surfaces). Besides the part of fracture energy consumed for determining the impact strength by Sharpie or Menazhe is spent on elastic and plastic (residual) specimen bending and tightening. This energy loss can be probably reduced by determining the impact toughness by Izod. Besides it has been mentioned that the maximal divergences between calculated and experimental values of $K_{Ic}(CT)$ are observed with the steels which, in process of specimens testing demonstrate the changes in their structure and properties, for example, with maraging steels with increased content if residual austenite.

It should be assumed that in case of long or lap (*crack*) seams when the relationship between the crack length $l$ and opening $a$ used in deriving equation (5) is disturbed, the criterion $K_{Ic}(CR)$ becomes invalid.

It conclusion let us once more stress the problems discussed in the work [1]. «Odd» dimension of $K_{Ic}[\text{Pam}^{1/2}]$ has probably appeared due to the fact that in equation (6) the member $a$ has been replaced by $\sqrt{\rho l}$, and equation (10), determining the dimension of $K_{Ic}$ has been obtained as a result of transferring $\sqrt{\rho}$ from the right-hand part of equation (9) into the left-hand one. In the mechanics of real-structure materials fracture these reexpressions have considerable significance because they allow to avoid the presence of the members $a$ and $\rho$ in equation (11), the values of which are impossible to find experimentally retaining in it the experimentally determined values $\sigma$ and $l$:

The sense of $K_{Ic}$ will be easier to understand if equation (I) is reduced to*:*

$$\sigma = K_{Ic} / c\sqrt{\pi l} \qquad (23)$$

From equation (23) it follows that $K_{1c}$ is a factor determining the dependence of ultimate stress limit ($\sigma_B$ or $\sigma_{theor.}$) of a material on the length of a crack (defect) present in it.

It should be supposed that in case of long or crooked cracks the relationship between the crack length $l$ and opening $a$ (which was used in generating equation **(5)**) is disturbed and the criterion $K_{1c}(TC)$ becomes invalid.

Conclusions:

1. The fundamental equation (5) of material fracture mechanics was obtained on the basis of Energy Conservation Law and without engaging linear fracture mechanics postulates.

2. From this equation it follows that the crack resistance $K_{1c}(CR) = 2\sigma_в \sqrt{l}$ and fracture toughness $G_{1c} = [K_{1c}(CR)]^2/2\sigma_в$ are respectively referred to as the force and energy criteria of non-linear fracture mechanics of an elastic-plastic material with real structure the tip radius of which $\rho > 0$, with the stress in this tip being $\sigma_y < \infty$.

3. The technique of obtaining the fundamental equation of material fracture mechanics excludes the necessity of applying the model of a real material transfer into quasi – brittle condition.

4. The dependence of $K_{1c}(CR)$ on $\sigma_в$, $\sigma_T$ and $KCU$ has been obtained and experimentally confirmed for a number of steels.

5. It is found feasible to verify the dependence found using other materials.

The Non-linear Fracture Mechanics (NLFM) is considered to comprise the basis of modern strength analysis of bearing structures and machine parts. The force criterion $K_{1c}(CR)=2\sigma_в\sqrt{l}$ and the energy criterion $G_{1c} = J_{1c} = (K_{1c})^2/2\sigma_в$ of material fracturing are referred to as the criteria of of NLFM.

**References.**

# Figures captions

Fig.1. Plate with thickness $\delta$ and central 2$l$-long and 2$a$-wide crack, loaded with $\sigma$; $\sigma_y$–stresses at the crack tips, $\sigma_y > \sigma$.

Fig.2. Pattern of the crack in Fig.1; $\Delta$ - minimal irretrievable crack divergence observed in its tips under the action of ultimate breaking stress ($\sigma_y - \sigma$).

Fig.3. Plate with thickness $\delta$ and lateral-long and $a$-wide crack

Figures:

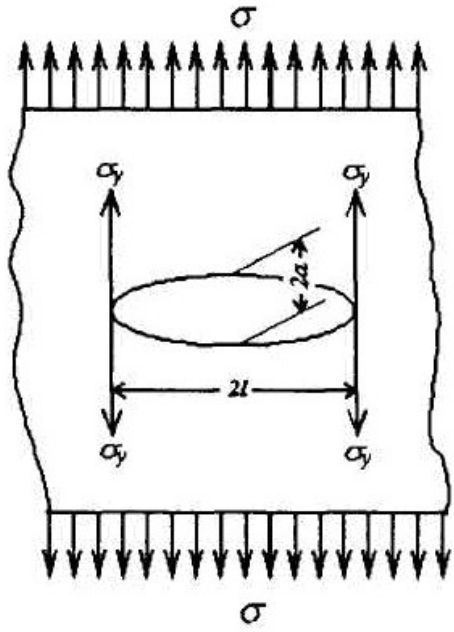

Fig.1

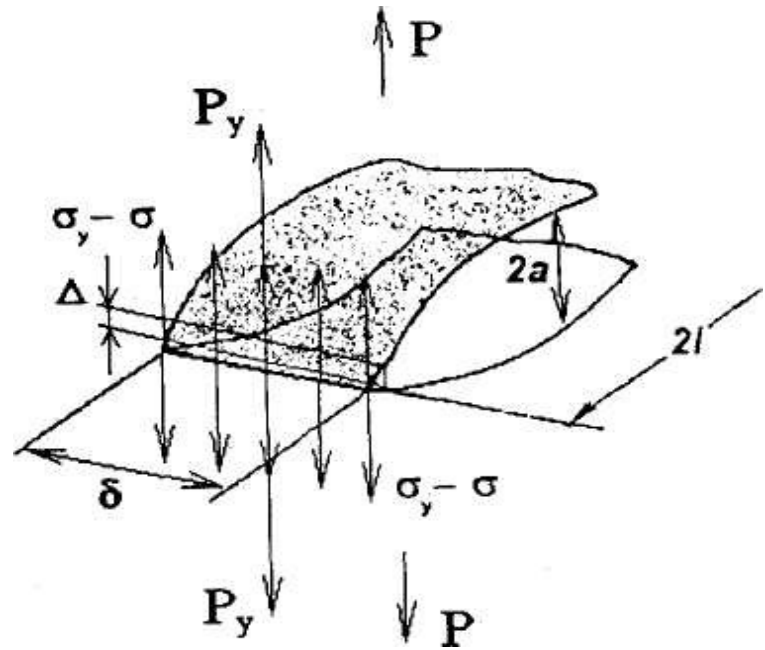

Fig.2

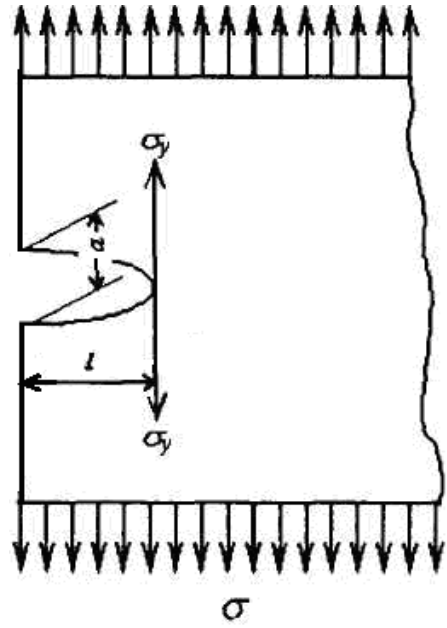

Fig.3